\documentclass[preprint,12pt]{elsarticle}

\usepackage{graphics}
\usepackage{graphicx}
\usepackage{epsfig}

\usepackage{amssymb}

\journal{Physics Letters B}

\begin{document}

\begin{frontmatter}



\title{Measurement of zero degree inclusive photon energy spectra for $\sqrt{s}=$ 900\,GeV proton-proton collisions at LHC}


%

\author[firenze-infn,firenze-univ]{O.\,Adriani}
\author[firenze-infn]{L.\,Bonechi}
\author[firenze-infn]{M.\,Bongi}
\author[firenze-infn]{G.\,Castellini}
\author[firenze-infn,firenze-univ]{R.\,D'Alessandro}
\author[nagoya]{K.\,Fukatsu}
\author[polyteque]{M.\,Haguenauer}
\author[nagoya]{T.\,Iso}
\author[nagoya,kmi]{Y.\,Itow\corref{cor1}}
\author[waseda]{K.\,Kasahara}
\author[nagoya]{K.\,Kawade}
\author[nagoya]{T.\,Mase}
\author[nagoya]{K.\,Masuda}
\author[firenze-infn,kmi]{H.\,Menjo}
\author[nagoya]{G.\,Mitsuka}
\author[nagoya]{Y.\,Muraki}
\author[catania-infn]{K.\,Noda}
\author[firenze-infn]{P.\,Papini}
\author[cern]{A.-L.\,Perrot}
\author[firenze-infn]{S.\,Ricciarini}
\author[nagoya,kmi]{T.\,Sako}
\author[nagoya]{K.\,Suzuki}
\author[waseda]{T.\,Suzuki}
\author[nagoya]{K.\,Taki}
\author[kanagawa]{T.\,Tamura}
\author[waseda]{S.\,Torii}
\author[catania-infn,catania-univ]{A.\,Tricomi}
\author[lbn]{W.\,C.\,Turner}

\address[firenze-infn]{INFN Section of Florence, Italy}
\address[firenze-univ]{University of Florence, Italy}
\address[nagoya]{Solar-Terrestrial Environment Laboratory, Nagoya University, Nagoya, Japan}
\address[kmi]{Kobayashi-Maskawa Institute for the Origin of Particles and the Universe, Nagoya University, Nagoya, Japan}
\address[polyteque]{Ecole-Polytechnique, Palaiseau, France}
\address[waseda]{RISE, Waseda University, Japan}
\address[cern]{CERN, Switzerland}
\address[kanagawa]{Kanagawa University, Japan}
\address[catania-infn]{INFN Section of Catania, Italy}
\address[catania-univ]{University of Catania, Italy}
\address[lbn]{LBNL, Berkeley, California, USA}

\cortext[cor1]{itow@stelab.nagoya-u.ac.jp}

\begin{abstract}

The inclusive photon energy spectra measured by the Large Hadron Collider forward (LHCf) 
experiment in the very forward region
of LHC proton-proton collisions at $\sqrt{s}=$ 900\,GeV are reported.
The results from the analysis of 0.30\,$\mathrm{nb^{-1}}$ of data collected
in May 2010 in the two pseudorapidity regions of $\eta\,>\,10.15$ and $8.77\,<\,\eta\,<\,9.46$
 are compared with the predictions of the hadronic interaction models DPMJET 3.04, EPOS 1.99, PYTHIA 8.145,
QGSJET I\hspace{-.1em}I-03 and SIBYLL 2.1, which  are widely used in ultra-high-energy cosmic-ray experiments. 
EPOS 1.99 and SYBILL 2.1 show a reasonable agreement with the spectral shape of the experimental data, 
whereas they predict lower cross-sections than the data.
The other models, DPMJET 3.04, QGSJET I\hspace{-.1em}I-03 and PYTHIA 8.145, are in good agreement with the data 
below 300\,GeV but predict harder energy spectra than the data above 300\,GeV. 
The results of these comparisons exhibited features similar to those for the previously reported data for 
$\sqrt{s}=$ 7\,TeV collisions.    

\end{abstract}

\begin{keyword}

LHC, Ultra-High Energy Cosmic Ray,  hadronic interaction models
\end{keyword}

\end{frontmatter}

\section{Introduction}
\label{Introduction}

The observations of ultra-high-energy cosmic rays (UHECR) have made notable improvements in the last few 
years\,\cite{ref-AugerSpectrum}\,\cite{ref-AugerAniso}\,\cite{ref-AugerComp}\,\cite{ref-HiResSpectrum}\,\cite{ref-HiResAniso}\,\cite{ref-HiResComp}\,\cite{ref-TA}.
However, although some critical parts of the interpretation rely on the Monte Carlo (MC) 
simulations of the air shower development, very forward particle emission in the hadronic interactions, 
which are relevant to the precise understanding of air showers,
have been poorly understood thus far, especially at such high energies.
To reduce the uncertainty in MC air shower simulations, the Large Hadron Collider 
forward (LHCf) experiment has performed measurements of the neutral  particles emitted 
to the very forward region of proton-proton collisions at the LHC.
In 2010, the operations at $\sqrt{s}=$\,7\,TeV and 900\,GeV were completed.
The photon energy spectra obtained from the 7\,TeV data have been previously reported elsewhere 
\,\cite{ref-7TeVphoton}.
Below this energy, the measurement of the $P_T$ spectra of $\pi^0$s by UA7 
in the rapidity region Y = 5.05 -- 6.65\,\cite{ref-UA7} is available. 
It is interesting to have a single-photon measurement at 0 degrees at different collision energies 
in order to discuss energy dependence. 
In this paper, we report the inclusive photon energy spectra in the very 
forward region for $\sqrt{s}=$\,900\,GeV proton-proton collisions 
with the same detectors and analysis methods as those used for the $\sqrt{s}=$\,7\,TeV analysis. 

Two LHCf detectors, called Arm1 and Arm2, were installed in the instrumentation slots 
of the TANs (Target Neutral Absorbers) located at $\pm$140\,m from the ATLAS interaction 
point (IP1) and  covering the pseudorapidity range from 8.7 to infinity (zero degrees). 
Each detector had two sampling and imaging calorimeters composed of 44 radiation lengths 
(1.55 hadron interaction lengths) of tungsten and 16 sampling layers of 3\,mm thick plastic scintillators.
The transverse sizes of the calorimeters were 20\,mm\,$\times$\,20\,mm and 40\,mm\,$\times$\,40\,mm in Arm1  
and 25\,mm\,$\times$\,25\,mm and 32\,mm\,$\times$\,32\,mm in Arm2.
The smaller and larger calorimeter of each Arm are called the small tower and large tower, respectively. 
The cross sections of the calorimeters, as observed from IP1, are illustrated in Fig.\,\ref{fig:detector}.
During the operations that were used in the analysis reported in this paper, 
a large fraction of the large Arm1 tower was obscured
from the IP as indicated by the shaded area in the Fig.\,\ref{fig:detector} due to the beam pipe 
material between the IP and the detector.

Four X-Y layers of position-sensitive detectors (scintillating fiber, SciFi, belts in Arm1 
and silicon micro-strip sensors in Arm2 with 1-mm and 0.16-mm readout pitches, respectively) were 
inserted to measure the transverse positions of the showers. 
The LHCf detectors have energy and position resolutions better 
than 5\,\% and 200\,$\mathrm{\mu m}$, respectively, for $>$100 GeV photons.
Detailed descriptions of the detectors can be found elsewhere\,\cite{ref-LHCfTDR}\,\cite{ref-LHCfJINST} 
\,\cite{ref-prototype}\,\cite{ref-sps2007}\,\cite{ref-LHCfsilicon}\,\cite{ref-menjopi0}.

This paper describes the first results of the analysis of inclusive photon energy 
spectra for $\sqrt{s}=$ 900\,GeV proton-proton collisions, which are primarily produced 
from the decay of $\pi^{0}$ and $\eta$ mesons generated in the collisions.
The data set and the MC simulation used in the analysis are introduced in Sec.\,\ref{Data} 
and  Sec.\,\ref{MCsimulation}, respectively. 
The analysis process is described in Sec.\,\ref{AnalysisandResult}. 
The experimental results and comparison with the MC predictions of several hadronic interaction models 
are presented in Sec.\,\ref{Result} and summarized in Sec.\,\ref{Summary}.

\section{Data}
\label{Data}

The data sets used in the analysis were taken on 2, 3 and 27 May 2010 during the LHC operations 
with proton-proton collisions at $\sqrt{s}=$ 900\,GeV, 
which correspond to the LHC fill identification numbers (Fill ID) 1068, 1069 and 1128, respectively.
In these fills, the LHC operated with one crossing bunch and one non-crossing bunch at IP1 
in Fill IDs 1068 and 1069 and with four crossing bunches and three non-crossing bunches at IP1 in Fill ID 1128.    
The luminosity ($L$) at IP1 during these fills was measured by the ATLAS experiment\,\cite{ref-luminosity}.
The luminosity during Fill ID 1068 and 1069 were  8 -- 3\,$\times$\,10$^{27}$\,$\mathrm{cm^{-2}s^{-1}}$ 
and 12 -- 4\,$\times$\,10$^{27}$\,$\mathrm{cm^{-2}s^{-1}}$, respectively.  
The total luminosity of the four crossing bunches in Fill ID 1128 was 
approximately 8\,$\times$\,10$^{27}$\,$\mathrm{cm^{-2}s^{-1}}$. 
The total integrated luminosity ($\int{L}dt$) during the LHCf operations in the three fills 
was 0.30\,$\mathrm{nb^{-1}}$. 
The uncertainty of the luminosity determination is $\pm\,21\,\%$\,\cite{ref-luminosity}.
The inelastic cross-section  ($\sigma_{inel}$) for a $\sqrt{s}=$\,900\,GeV proton-proton collision 
was estimated to be $\sigma_{inel}=$\,53.0\,mb
from the predictions of the total cross-section and  the elastic cross-section, which are based 
on the recent experimental results\,\cite{ref-TOTEM}\,\cite{ref-PDG}. 
The number of inelastic collisions ($N_{inel}$) during the three fills was calculated to be $1.58\times10^{7}$.
The luminosity during the three fills is summarized in Tab.\ref{tab-lumi}.

During the $\sqrt{s}=$\,900\,GeV fills, the LHCf operations were performed with a high-gain operation of the PMTs  
for the sampling layers to detect photons with energies as low as 50\,GeV with a nearly 100\,\% 
trigger efficiency and with a lower threshold with respect to the $\sqrt{s}=$\,7\,TeV data. 
The typical PMT gain for the high-gain operations was 3 -- 5 times higher than the nominal gain 
that was used to obtain the 7 TeV data. 
Neither the saturation of PMTs nor the range of ADC caused problems    
because the maximum energy of the incident photons were expected to be 8 times lower than 
those for the 7 TeV data. 
The data acquisition (DAQ) triggers were generated from beam pickup signals (BPTX) 
followed by ``shower trigger'' signals. 
The trigger condition of the shower trigger was that signals from any three successive 
scintillator layers in any calorimeters exceed 
the predefined threshold (approximately 17 MeV for high gain).
The average DAQ live times during the LHCf $\sqrt{s}=$\,900\,GeV operations was 99.2\,$\%$ (Arm1) 
and 98.0\,$\%$ (Arm2). 
The total numbers of triggered events in Arm1 and Arm2 were 44,389 and 62,916, respectively.
Because of the very low luminosity and the low event rate per inelastic collision, 
the probability of the pile-up of events was $<\,10^{-4}$, negligibly small. 

\begin{table*}[t]
	\begin{center}
	\begin{tabular}{llrrcc}
	\hline
	Date & Time\,(UT) & Fill ID & $L$ ($\mathrm{cm^{-2}s^{-1}}$)  & $\int{L}dt$  & Crossing bunch\\
	\hline
	2 May 2012   & 12:50 -- 19:23 & 1068 &   8\,$\sim$\,3\,$\times$\,10$^{27}$ & 0.11 $\mathrm{nb^{-1}}$ & 1$\times$1\\
	3 May 2012   & 00:17 -- 07:08 & 1069 & 12\,$\sim$\,4\,$\times$\,10$^{27}$ & 0.17 $\mathrm{nb^{-1}}$ & 1$\times$1\\
	27 May 2012 & 13:18 -- 14:03 & 1128 &       8\,$\times$\,10$^{27}$ & 0.02  $\mathrm{nb^{-1}}$ & 4$\times$4\\
	\hline
	\end{tabular}
	\caption{
	  Summary of the luminosity during  $\sqrt{s}=$ 900\,GeV operations\,\cite{ref-luminosity}. }
	\label{tab-lumi}
	\end{center}
\end{table*}

\section{MC simulation}
\label{MCsimulation}

To compare the experimental results with the predictions of hadronic interaction models, 
MC simulations were performed with the hadronic interaction models, 
QGSJET I\hspace{-.1em}I-03\,\cite{ref-QGS2}, PYTHIA 8.145\,\cite{ref-PYTHIA8a}\,\cite{ref-PYTHIA8b}, SIBYLL 2.1\,\cite{ref-SIBYLL}, 
EPOS 1.99\,\cite{ref-EPOS} and DPMJET 3.04\,\cite{ref-DPM3}. 
In the MC simulations, $3\,\times10^{7}$ inelastic proton-proton collisions were generated by each model, 
and the secondaries were transported in the beam pipe from IP1 to the LHCf detectors.  
The magnetic fields of the dipole magnets 
located between IP1 and the LHCf detectors were taken into account.  
The detector response was calculated using the EPICS 8.81/COSMOS 7.49 simulation package\,\cite{ref-EPICS}.

In addition, $1.0\,\times10^{8}$ events were generated using QGSJET I\hspace{-.1em}I-03. This data set 
was used for studies of the detector response and particle identification (PID) correction described 
in Sec.\,\ref{PIDselection}. 

\section{Analysis}
\label{AnalysisandResult}
\subsection{Energy Reconstruction}
\label{EnergyReconstruction}

The sum of the energy deposited in the 2$^{nd}$ to 13$^{th}$ scintillator layers, 
after corrections for gain variation and the non-uniformity of the light yield of each scintillator layer, 
was used as an energy estimator for the primary photons incident on the LHCf detectors. 
Each PMT gain was premeasured for various HVs using N$_2$ laser calibration. 
The calibration of the deposited energy in each layer was performed for different 
HV settings (low-, nominal-, and high-gain operations) using 50 -- 200\,GeV/$c$ electron beams 
and 150\,GeV/$c$ muon beams at the CERN SPS\,\cite{ref-sps2007}. 
The non-uniformity of the light yield of each scintillator layer was 
measured using a $\beta$-ray source before assembling the detectors.  
Because a fraction of the shower particles leak out of the sides of the calorimeters (`shower leakage'), 
the total energy deposited was corrected for `shower leakage' by a function of the shower impact position.
This function was determined by the MC simulation. 
The impact positions of the showers were determined using the information from the position-sensitive layers. 
The events that fell within 2\,mm of the edges of calorimeters were removed from the analysis 
to avoid the degradation of the energy resolution due to `shower leakage'.
We set the energy threshold of this analysis to 50\,GeV to avoid background from the interactions 
between secondary particles and the beam pipe,
which was expected to be concentrated below 50\,GeV according to a MC simulation.
The trigger efficiencies of both Arms were also checked by two samples. 
One was an unbiased data sample with nominal gain triggered by the shower triggers 
at the opposite side of the detector. 
The other was the detector MC simulation for both high and nominal gains. 
Considering the difference of gains (a factor 3 -- 5) between the high and nominal gain operations,   
the two methods gave consistent results. 
We found that the efficiencies were 100\% for $>$\,30\,GeV 
incident photons  for both arms for the higher gain. 
This was sufficiently lower than the 50 GeV analysis threshold. 

Similar to the previous 7\,TeV analysis, the systematic uncertainties of 
the absolute energy scale were evaluated from the reconstructed invariant 
mass of the $\pi^0$s from the 7\,TeV data taken with a nominal gain.
Additionally, we checked $\pi^0$ mass peaks in the 7\,TeV data with a higher gain 
taken in a different period. We found +2.7\,\% (+0.7\,\%) differences of the mass 
peaks from those for the nominal gain in Arm1 (Arm2). 
These differences were compatible with the uncorrelated energy scale errors ($\pm$3.5\,\%)  
quoted for the energy scale calibration of the detectors using 
the SPS beams or a long-term time variation. 
Conservatively, they were added to the energy scale's systematic error in quadrature.
Finally, the energy scale uncertainties were estimated to be 
[$-$10.2\,\%,\,$+$1.8\,\%] and [$-$6.6\,\%,\,$+$2.2\,\%] for Arm1 and Arm2, respectively.
The systematic uncertainties of the energy spectra due to the energy scale uncertainty are 
listed for the Arm1 and the Arm2 detectors in Tab.\,\ref{tab-sysene1} and Tab.\,\ref{tab-sysene2}, 
respectively. 

\begin{table*}[t]
	\begin{center}
	\begin{tabular}{|l|c|c|c|c|c|}
	\hline
	Energy range (GeV)  & $50$ -- $150$  & $150$ -- $200$ & $200$ -- $250$ & $250$ -- $300$ & $300$ -- $450$ \\
	\hline
	Arm1 Small (\%)     &  $-2$,\,$+3$  & $-6$,\,$+3$   & $-26$,\,$+5$  & $-48$,\,$+14$ & $-48$,\,$+14$  \\
        \hline
	Arm1 Large (\%)     & $-8$,\,$+5$  & $-20$,\,$+5$   & $-20$,\,$+5$  & $-50$,\,$+15$ & $-71$,\,$+30$  \\
	\hline
	\end{tabular}
	\caption{
	  Summary of the systematic errors for the energy spectra of Arm1 due to the energy scale uncertainty. }
	\label{tab-sysene1}
	\end{center}
\end{table*}

\begin{table*}[t]
	\begin{center}
	\begin{tabular}{|l|c|c|c|c|}
	\hline
	Energy range (GeV)  & $50$ -- $150$  & $150$ -- $225$ & $225$ -- $300$ & $300$ -- $450$  \\
	\hline
	Arm2 Small (\%)     &  $-15$,\,$+8$  & $-15$,\,$+8$   & $-28$,\,$+12$  & $-66$,\,$+34$ \\
        \hline
	Arm2 Large (\%)     & $-5$,\,$+7$  & $-20$,\,$+7$   & $-24$,\,$+10$  & $-63$,\,$+14$  \\
	\hline
	\end{tabular}
	\caption{
	  Summary of the systematic errors for the energy spectra of Arm2 due to the energy scale uncertainty. }
	\label{tab-sysene2}
	\end{center}
\end{table*}

The energy of photons in multi-hit events with more than one photons incident on a single tower 
would not be reconstructed correctly. 
We estimated the possible bias in the energy reconstruction due to double-incident events  
using MC simulation using QGSJET I\hspace{-.1em}I-03. We found that the number of events whose true energies  
were modified by more than 2\,\% was very small. The fraction of such events 
was expected to be less than 1\,\% of the events having a total incident energy of $>$40 GeV.
Although such multi-hit events can be identified using the lateral distributions measured 
by the position-sensitive layers,
we did not apply the multi-hit cut or any correction to the spectra 
in this analysis to avoid a bias due to the misidentification 
of single photon events as multi-hit events. 
The effect was at most 1\% in the lower energy bins.

\subsection{Photon Event Selection}
\label{PIDselection}
\begin{figure}
	\centering
	\includegraphics[width=11cm,clip]{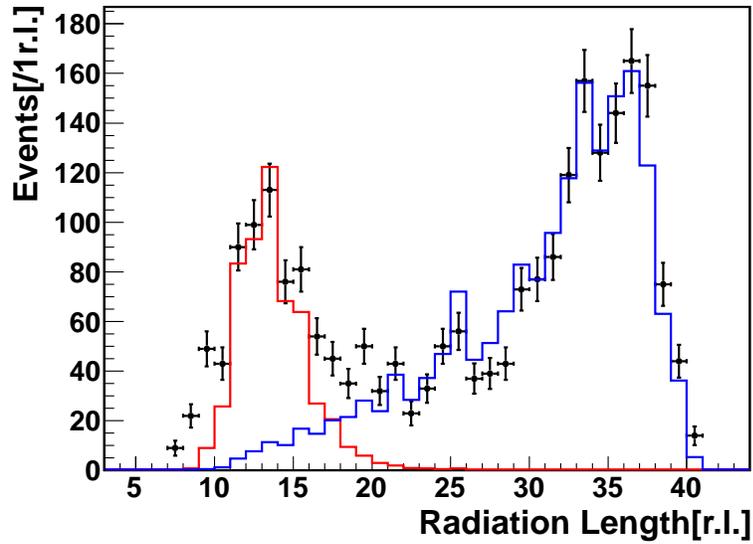}
	\caption{
	The $L_{90\%}$ distribution measured by the Arm1 small tower for the reconstructed 
        energy range of 50 -- 100\,GeV.
	The black points show the experimental data; the red and blue histograms are the $L_{90\%}$ 
        distributions of MC calculation for pure photons and pure hadrons (`templates'), respectively.
	These template histograms are independently normalized to give the best match to the experimental result.
	}
	\label{fig:L90distribution}
\end{figure}

To select the electromagnetic shower events and to eliminate contamination by hadronic shower events, 
a parameter called $L_{90\%}$ was defined.
$L_{90\%}$ is a longitudinal length in units of radiation length\,(r.l.) in which 90\,$\%$ 
of the total shower energy is deposited in the calorimeter.
Figure\,\ref{fig:L90distribution} shows the $L_{90\%}$ distribution of the 
Arm1 small tower events with reconstructed energy in the range of 50 to 100\,GeV.
The two peaks near 13 r.l. and 35 r.l. correspond to electromagnetic showers and hadronic showers, respectively.
Fig.\,\ref{fig:L90distribution} also shows the $L_{90\%}$ distributions generated by the MC simulation 
with QGSJET I\hspace{-.1em}I-03 for pure photons and pure hadrons (neutrons). 
These MC distributions have been normalized to the $L_{90\%}$ distribution of the experimental data.  
They are hereafter called the `template'.
We set the $L_{90\%}$ criteria to keep the photon selection efficiency $\epsilon_{PID}$\,=\,90$\%$ 
over the entire energy range based on the template for photons.
In Fig.\,\ref{fig:L90distribution} this would correspond to $L_{90\%}=$\,16.8 r.l.

The purity ($\mathcal{P}$)  of a photon sample was estimated by normalizing the templates for photons 
and for hadrons to the measured $L_{90\%}$ distribution 
 for each energy range (`template fitting' of the $L_{90\%}$ distribution). 
 $\mathcal{P}$ was defined as $\mathcal{P}=N_{photon}/(N_{photon}+N_{hadron})$ in each energy region.
Here $N_{photon}$ and $N_{hadron}$ are the numbers of photons and hadrons in the selected $L_{90\%}$ range 
in the template, respectively.
The correction factor $\mathcal{P}\times\epsilon^{-1}_{PID}$ was applied to the number of events in each energy 
bin to correct for the  inefficiency of the photon selection and  for the residual contamination by hadrons.

However, there were small discrepancies between the $L_{90\%}$ distributions of the experimental 
data and of the MC simulations. 
These discrepancies may be caused by errors in the absolute energy determination or 
in the channel-to-channel gain calibrations.
Here, we consider the systematic uncertainty caused by the uncertainty of the `template fitting' 
method for obtaining photon spectra. 
Small modifications of the template  
(widening with respect to the peak position up to 30\,$\%$ and 
a constant shift up to 1.0 r.l. for Arm1 and 0.8 r.l. for Arm2,
to give the best match with the data) provide the size of uncertainty in the correction 
factors to the photon spectra. 
The difference of the correction factors between the original and the modified template methods amounted 
to $\pm$10\,\% ($\pm$12\,\%) and $\pm$55\,\% ($\pm$44\,\%) for below and above 150\,GeV photon energy, respectively,  
in the Arm1 small (large) tower.
The difference of the correction factors were $\pm$20\,\% ($\pm$25\,\%) 
in the Arm2 small (large) tower for entire photon energy. 
These numbers were assigned as the systematic uncertainty of the energy spectra due to PID errors.

\subsection{Background Subtraction}
\label{BGandUncertainty}
The background particles from the interactions between the proton beams and the residual gas 
in the vacuum beam pipe hit the detectors synchronously with the beam-beam events.
The amount of this background can be estimated using the events triggered by the passage of non-crossing bunches.
Assuming that the beam intensity of each bunch was same, 
the background levels were estimated as approximately 1\,\% and 2\,\% for the small and large towers, respectively, 
in both Arms. The estimated backgrounds were subtracted from the energy spectra.

\subsection{Beam Center Position}

The projected position of the zero-degree collision angle at the LHCf detectors, referred 
as the `beam center', is  an important parameter in the geometrical analysis of the experimental data.  
Because the flux of the secondary particles produced by $\sqrt{s}$\,=\,900\,GeV proton-proton collisions 
was expected to be uniform over the acceptance of the LHCf detectors, the beam center could not be determined 
directly from our measurements. 
In this analysis, we assumed that the beam center was at the center position determined by the alignment  
survey of the detectors.
The beam center was located near the center of the small calorimeter of each Arm as shown in Fig.\,\ref{fig:detector}.
The beam position and the beam angle were monitored by the Beam Position Monitor (BPMSW) 
installed $\pm$\,21\,m from IP1\,\cite{ref-BPM}.   
The fill-by-fill fluctuation of the calculated `beam center' at the LHCf detectors during 
the period for the presented data set was approximately 4\,mm. 
We assigned a systematic uncertainty of $\pm$2\,mm to the beam center. 
To estimate the effect of this uncertainty  on the energy spectra, we defined an area that 
had a slightly narrower acceptance than the calorimeter, and an energy spectrum was generated 
from the events falling within this area.
Additionally, four spectra were made by shifting the area by $\pm$2\,mm vertically and horizontally.
As a systematic uncertainty  of energy spectra due to the uncertainty of the beam center,  
we assigned the differences of the spectra as shown in Tab.\,\ref{tab-syscen1} and Tab.\,\ref{tab-syscen2} 
for the Arm1 and the Arm2 detectors, respectively. 

\begin{table*}[t]
	\begin{center}
	\begin{tabular}{|l|c|c|c|}
	\hline
	Energy range (GeV)  & $50$ -- $150$  & $150$ -- $250$ & $250$ -- $450$  \\
	\hline
	Arm1 Small (\%)     &  $-7$,\,$+2$  & $-7$,\,$+2$   & $-17$,\,$+2$   \\
        \hline
	Arm1 Large (\%)     & $-11$,\,$+6$  & $-11$,\,$+12$   & $-15$,\,$+12$  \\
	\hline
	\end{tabular}
	\caption{
	  Summary of the systematic errors for the energy spectra of Arm1 due to the beam center uncertainty. }
	\label{tab-syscen1}
	\end{center}
\end{table*}

\begin{table*}[t]
	\begin{center}
	\begin{tabular}{|l|c|c|c|}
	\hline
	Energy range (GeV)  & $50$ -- $175$  & $175$ -- $250$ & $250$ -- $450$  \\
	\hline
	Arm2 Small (\%)     &  $-3$,\,$+3$  & $-4$,\,$+9$   & $-4$,\,$+13$  \\
        \hline
	Arm2 Large (\%)     & $-3$,\,$+4$  & $-3$,\,$+6$   & $-16$,\,$+14$  \\
	\hline
	\end{tabular}
	\caption{
	  Summary of the systematic errors for the energy spectra of Arm2 due to the beam center uncertainty. }
	\label{tab-syscen2}
	\end{center}
\end{table*}

\section{Energy Spectra Results}
\label{Result}
\subsection{Reconstruction of Energy Spectra}

\begin{figure}
	\centering
	\includegraphics[width=12cm,clip]{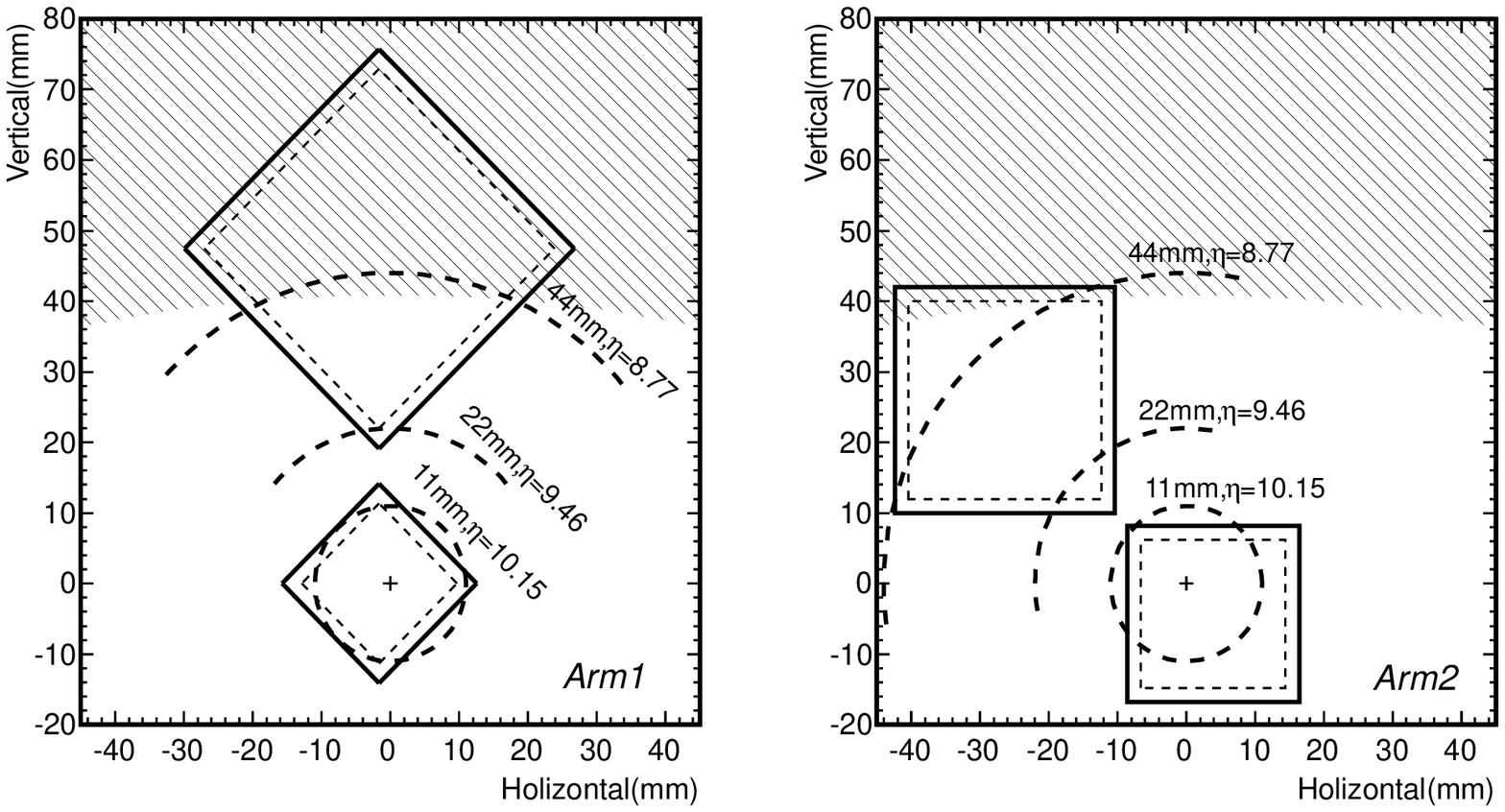}
	\caption{
	The cross-sections of the calorimeters viewed from IP1, left for Arm1 and right for Arm2. 
	The cross marks on the small calorimeters indicate the projections of the zero-degree collision 
        angle onto the detectors (`beam center').
	 The shaded areas in the upper parts of the figure indicate the shadows of the beam pipes 
         located between IP1 and the detectors, where 
	the detectors are insensitive to the detection of IP1 proton-proton collision products.  
	The dashed squares indicate the border of a 2\,mm edge cut described in Sec.\,\ref{EnergyReconstruction}.
	The circles and the arcs indicate the distance ($r$) from the beam center of 11\,mm, 22\,mm and 44\,mm, 
	and the pseudorapidity ($\eta$) of 10.15, 9.46 and  8.77, respectively. 
	 In this analysis,  the events in the regions of $r\,<$\,11\,mm and 22\,mm\,$<\,r\,<$\,44\,mm were used.  
	 }
	\label{fig:detector}
\end{figure}

\begin{figure}
	\centering
	\includegraphics[width=14cm,clip]{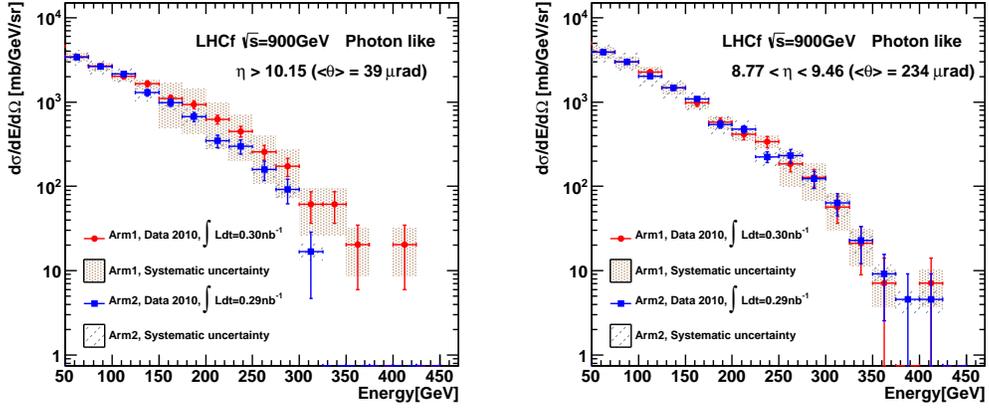}
	\caption{
	Photon energy spectra measured by Arm1 (red circles) and Arm2 (blue rectangles) in units of  differential 
        cross-section $d\sigma/dEd\Omega$.
	The left and the right panels show the results of the small towers and the large towers, respectively.
	The pseudorapidity coverages of the small and the large towers are $\eta\,>\,10.15$ 
        and $8.77\,<\,\eta\,<\,9.46$, respectively.
	Considering the geometrical shape of the calorimeters,  the averages of the polar angle $\langle\theta\rangle$ with respect to the beam axis are 39\,$\mu$rad and 234\,$\mu$rad for the small and the large towers, respectively.
	The error bars indicate the statistical errors, and the hatched areas indicate the total systematic uncertainties that come from particle identification and the `beam  center'  position.
	}
	\label{fig:spectra_arm12}
\end{figure}

To reduce a possible pseudorapidity ($\eta$) dependence when comparing and 
combining the energy spectra measured by the two Arms, we selected Arm2 events 
with a pseudorapidity range similar to that of Arm1.
For the small tower, we selected events with the distance ($r$) from the beam center less than 11\,mm, 
which corresponded to the pseudorapidity range of $\eta\,>\,10.15$ (the circles in Fig.\,\ref{fig:detector}).
Similarly, for the large tower, we set the conditions as 22\,mm\,$<\,r\,<$\,44 mm, which corresponded to 
the pseudorapidity range of $8.77\,<\,\eta\,<\,9.46$ (the arcs in Fig.\,\ref{fig:detector}).
The calorimeters did not uniformly cover the pseudorapidity ranges as shown in Fig.\,\ref{fig:detector}.
We confirmed that there was a negligible pseudorapidity dependence of the energy spectra inside  
each pseudorapidity range. 
The reconstructed photon energy spectra of Arm1 and Arm2 are shown in Fig.\,\ref{fig:spectra_arm12} 
in units of differential cross-sections $d\sigma/dEd\Omega$, where $E$ is the photon energy  
and $\Omega$ is the solid angle. 
The differential cross-section was calculated as $d\sigma/dEd\Omega = 1/(\int{L}dt) \; dN/dEd(cos\theta)d\phi$, 
where $N$ is the number of events in each energy bin,  
$\int{L}dt$ is the integrated luminosity after the correction for the DAQ live time for each Arm and 
$\theta$ and $\phi$ are the polar and the azimuthal angles with respect to the beam axis, respectively. 
Considering the geometrical acceptance of the calorimeters, 
the averages of the polar angle $\langle\theta\rangle$  are 39\,$\mu$rad and 234\,$\mu$rad 
for the small and the large tower, respectively. 
The error bars in Fig.\,\ref{fig:spectra_arm12}  indicate the statistical uncertainty;  
the hatched areas show the systematic errors in which the particle identification 
and the beam position uncertainties were taken into account. 
Because the systematic uncertainties  due to the energy determination may be correlated between 
Arm1 and Arm2\,\cite{ref-7TeVphoton},  they were not taken into account in Fig.\,\ref{fig:spectra_arm12}. 
The two spectra from Arm1 and Arm2 in each pseudorapidity region gave consistent results 
within the statistical and the systematic errors. 

\begin{figure}
	\centering
	\includegraphics[width=14cm,clip]{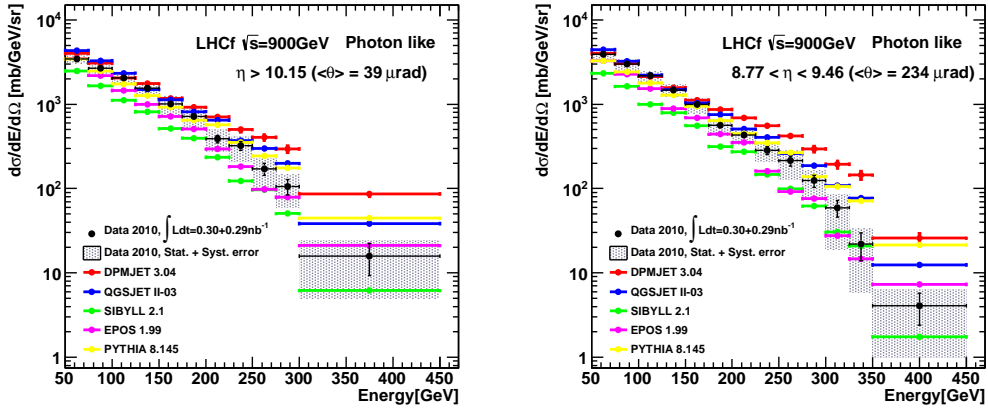}
	\caption{
	Combined Arm1 and Arm2 photon energy spectra compared with MC predictions.
	The data from Arm1 and Arm2 correspond to the integral luminosities of 0.30 and 0.29\,nb$^{-1}$, respectively.
	The left and the right panels are the results of the small ($\eta\,>\,10.15$) 
        and the large ($8.77\,<\,\eta\,<\,9.46$) towers, respectively.
	The black points indicate the experimental data with the statistical uncertainty (error bars) and  the total uncertainty,   quadratical summation of the statistical and the systematic errors (black hatches).
	The systematic uncertainty of the luminosity determination ($\pm21\,\%$) is not taken into account in the errors. 
	The colored points indicate the results of MC predictions, QGSJET I\hspace{-.1em}I-03 (blue), PYTHIA 8.145 (yellow), SIBYLL 2.1 (green), EPOS 1.99 (magenta) and DPMJET 3.04 (red). 
	Only the statistical uncertainty of DPMJET 3.04 is shown by the error bars  as representative of the models. 
	}
	\label{fig:spectra_comparison}
\end{figure}

The combined energy spectra of Arm1 and Arm2 are shown in Fig.\,\ref{fig:spectra_comparison} as weighted averages,
with the weights taken to be the square of the inverse of the errors  in each energy bin.  
The error bars of the data (black points) represent the statistical error;  
the hatches in the spectra represent the total uncertainty  (quadratical summation of the statistical 
and the systematic errors).
The sources of the systematic error are the particle identification and 
the beam position uncertainties. The energy scale errors were also included,  
assuming a correlation between the two Arms. 
Note that the uncertainty of the luminosity determination ($\pm21$\,\%) is not shown 
in Fig.\,\ref{fig:spectra_comparison}.
It can introduce a constant vertical shift of the spectra, but it cannot change the shapes of the spectra.   
We see a smooth spectrum from each of the two pseudorapidity regions, considering the errors. 
The similarity of the two spectra suggests only a small pseudorapidity dependence 
between the two pseudorapidity regions. 

\subsection{Comparison with Models}
\label{SpectraComp}

\begin{figure}
	\centering
	\includegraphics[width=14cm,clip]{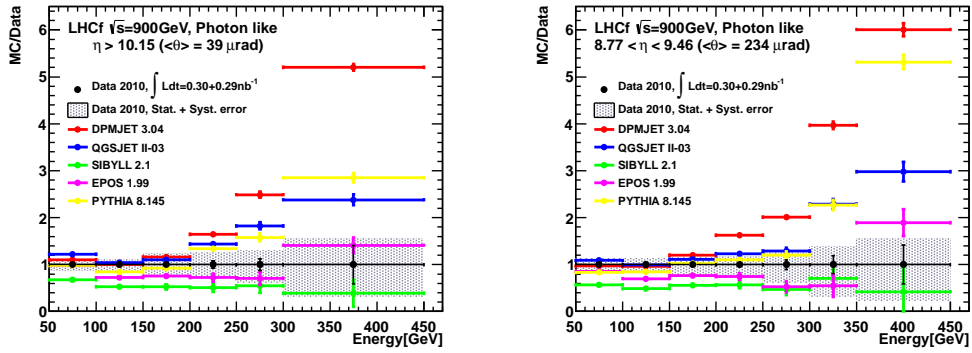}
	\caption{
	Ratio of the MC spectra divided by the data in each energy bin. 
	The left and the right panels show the spectra for the pseudorapidity ranges of  $\eta\,>\,10.15$ and $8.77\,<\,\eta\,<\,9.46$, respectively.
	The colored plots indicate the results of MC, QGSJET I\hspace{-.1em}I-03 (blue), PYTHIA 8.145 (yellow), SIBYLL 2.1 (green), EPOS 1.99 (magenta) and DPMJET 3.04 (red). 
	 To describe the size of the errors, the experimental data are also shown on the ratio of unity with the statistical uncertainty (error bars) and the total uncertainty of data (the black hatches).  
The luminosity uncertainty was not included. 
	The statistical uncertainties of each MC are shown as the error bars of MC.  
	}
	\label{fig:spectra_ratio}
\end{figure}

In Fig.\,\ref{fig:spectra_comparison}, the predictions of the hadronic interaction models,  
QGSJET I\hspace{-.1em}I-03, PYTHIA 8.145, SIBYLL 2.1, EPOS 1.99 and DPMJET 3.04, are also shown.
The same analysis processes were applied to the MC simulations as to the experimental data except 
for the particle identification using $L_{90\%}$ and its correction.  
For the analysis of the MC simulations, the known particle type was used. 
For better visibility,  only the statistical errors for DPMJET 3.04 (red points) are shown by the error bars.
Figure\,\ref{fig:spectra_ratio} shows the ratios of the MC spectra divided by the data in each energy bin.  
In Fig.\,\ref{fig:spectra_ratio},  the statistical error of each MC is shown as the error bar of each point.
The trends of the experimental data compared to each MC are similar for the two pseudorapidity ranges. 
EPOS 1.99 and SYBILL 2.1 show a reasonable agreement with the spectral shape of the experimental data, 
whereas they predict lower cross-sections than the data.
The other models, DPMJET 3.04, QGSJET I\hspace{-.1em}I-03 and PYTHIA 8.145, are in good agreement with the data 
below 300\,GeV but predict harder energy spectra than the data above 300\,GeV. 
The trends of the experimental data compared to the MC predictions in Fig.\,\ref{fig:spectra_ratio} 
are similar to those for the single-photon energy spectra in the pseudorapidity 
$\eta > 10.94$ previously reported for $\sqrt{s}=7$TeV proton-proton collisions\,\cite{ref-7TeVphoton}. 

\section{Summary}
\label{Summary}

LHCf measured the forward inclusive photon energy spectra for $\sqrt{s}=$\\900\,GeV 
proton-proton collisions in May 2010. 
The total integrated luminosity of the data set used in this analysis is 0.30 nb$^{-1}$.
The two LHCf detectors (Arm1 and Arm2) gave consistent results within the statistical 
and systematic errors for the small and the large towers, which cover the pseudorapidity 
ranges of  $\eta\,>\,10.15$ and  $8.77\,<\,\eta\,<\,9.46$, respectively.  
The combined energy spectra of Arm1 and Arm2 were compared with the predictions of 
five hadronic interaction models,  DPMJET 3.04, EPOS 1.99, PYTHIA 8.145, QGSJET I\hspace{-.1em}I-03 and SIBYLL 2.1. 
EPOS 1.99 and SIBYLL 2.1 reproduce well the shape of the experimental energy spectra, 
but they predict a lower cross-section than the LHCf data. 
The other models predict harder spectra than the LHCf data above 300\,GeV.  
These results of comparison exhibited features similar to those for the previously 
reported data for $\sqrt{s}=$ 7\,TeV collisions.    

\section*{Acknowledgements}
We thank the CERN staff and the ATLAS collaboration for their essential contributions 
to the successful operation of LHCf. 
This work is partly supported by Grant-in-Aid for Scientific Research by MEXT of Japan, 
the Mitsubishi Foundation in Japan and INFN in Italy. The receipts of JSPS
Research Fellowship (HM and TM), INFN fellowship for non-Italian citizens
(HM and KN) and the GCOE Program of Nagoya University `QFPU' from
JSPS and MEXT of Japan (GM) are also acknowledged. 
A part of this work was performed using the computer resource provided 
by the Institute for the Cosmic-Ray Research (ICRR), University of Tokyo.





\bibliographystyle{model1-num-names}
\bibliography{<your-bib-database>}



\clearpage

\end{document}